\documentstyle[amscd,amssymb,12pt]{amsart}

\def\hcorrection#1{\advance\hoffset by #1 }
\def\vcorrection#1{\advance\voffset by #1 }
\vcorrection{-.5in}
\hcorrection{-.7in}

\input xypic \xyoption{curve}
\input epsf

\begin{document}

\title{Quantum Relativity}         
\author{Lucian M. Ionescu}        
\date{May 15, 2010}          
\begin{abstract}
Quantum Relativity is supposed to be a new theory,
which locally is a deformation of Special Relativity,
and globally it is a background independent theory including the main ideas
of General Relativity, with hindsight from Quantum Theory.

The qubit viewed as a Hopf monopole bundle
is considered as a unifying gauge ``group''.
Breaking its chiral symmetry is conjectured to yield
gravity as a deformation of electromagnetism.

It is already a quantum theory in the context of Quantum Information Dynamics
as a discrete, background independent theory,
unifying classical and quantum physics.

Based on the above, Quantum Gravity is sketched as an open project.
\end{abstract}

\maketitle
\tableofcontents

\section{Preamble}
The present article aims to present two important ideas in the context
of a general conceptual framework called {\em Quantum Relativity},
which is the ``space-time oriented'' companion of Quantum Information Dynamics:

\vspace{.1in}
\quad 1) The unifying gauge ``group'' is the qubit, viewed as the {\em Hopf monopole bundle}:
$$SU(1)\cong S^1\to SU(2)\cong S^3\to S^2;$$

\quad 2) ``Gravity'' is a {\em deformation} of Electromagnetism.

\vspace{.1in}
Therefore in the context of the discrete background independent Quantum Information
Dynamics, it is already a ``quantum'' theory.

\vspace{.1in}
The focus will be on the physics interface, 
supported by mathematical ideas from deformation theory.

There will be no details nor final formulas, since the present approach is tentative at this stage,
occasionally presenting conflicting alternatives as part of the 
quantum gravity ``puzzle'' ... 

Instead, the author hopes that the reader will be interested in pursuing some of the
questions raised directly or, quite often indirectly, through claims based on intuition.

\section{Introduction}       
From a Computer Science point of view of Mathematical-Physics
a new paradigm emerges: {\em Quantum Information Dynamics} \cite{I-MPCS}.
It is the mathematical-physics component of the {\em Digital World Theory} \cite{I-DWT},
providing a unifying framework for both classical and quantum physics,
based on graph complexes and their cohomology \cite{Kon,CK,I-FL,IF}.

The main point is that the role of Electromagnetism was underestimated, even 
after the remarkable discovery of Aharonov-Bohm a. a. of the close connection
between quantum phase and {\em classical} EM as a $SU_1$-gauge theory.

A closer inspection of {\em Special Relativity} reveals that not only time is not ``absolute''
as a conclusion of a Einstein's critical analysis of the concept of synchronization,
but also the concept of ``direction'' (parallelism) is subject to the same criticism.
Therefore an $SO_3$-connection is mandatory to make sense of ``direction correlation'';
the assumption of a metric with its induced Levi-Civita connection is clearly
an extra assumption, no longer appropriate in a modern physics dominated
by quantum theory.

In conclusion \cite{I-MPCS}, EM as a $SU_1$-gauge theory is only a Hopf ``fiber'' of 
QID as a $SU_2$-Yang-Mills theory on whatever ``space'' is.
The author claims that ``space'' should be a model for the structure of matter,
preferably a dynamical lattice, as argued in \cite{I-DWT},
and ``time'' per se, with physical meaning is not one dimensional,
since it should model the {\em change of structure}, preferably as a cobordism
\footnote{Categorical approach: systems/objects, processes/morphism.}.

This yields a nice {\em Background Space-Time Independent Theory} (BIT),
from which Quantum Mechanics (QM) emerges, 
with Heisenberg ``uncertainty relations'' as its trademark due to 
an averaging over possible space-time coordinate systems, 
which are embedding in an ambient space-time manifold as in {\em String Theory}.

From QID also {\em Quantum Field Theory} emerges, with its Feynman diagrams and variable 
number of particles trademark, when taking full advantage of graph homology 
and loop derivative, as in {\em Loop Quantum Gravity}.

When we say ``emerges'', we mean that a reasonably simple {\em interface} exists between the
two theories (conceptual level), allowing to incorporate the results of one theory into the other,
as if it would be a ``plug-in'', once the differences of languages used for their
implementations are identified.

But then, why is there a ``Gravity'' to ``spoil'' this nice picture?

In this article we will (try) to prove that ``Gravity'' is a perturbation of
``Electromagnetism'' (EM), related to the standard bialgebra deformation technique
from quantization
\footnote{Quotations are in order since there are also various versions 
of EM.}. 

Although the ``bigger picture'' seems to indicate that the ``speed of light'' is 
in fact an isomorphism (with four eigenvalues) corresponding to a 
family of deformations of Hodge duality,
probably dual to Dirac quantization at the level of infinitesimal symmetries:
$$F(G)_c \times U(g)_h\to \mathbf{C},$$
we will restrict our attention to {\em deforming Special Relativity}.

Indeed, Lorentz group is a one-parameter {\em infinitesimal} deformation of Galilei group.
In view of General Relativity, it represents the local symmetries of physical processes
(classical or quantum).
Therefore we believe that one needs to \underline{deform} Galilei group,
{\em not} to ``quantize'' General Relativity!
As hinted above, we claim that ``quantization'' comes for free, 
from the discretization process and categorical approach
\footnote{What ``is'' and its ``change'' do not commute as in pointwise physics.}.

The byproduct of deformation of symmetries, via Noether Theorem and Euler-Lagrange Equations,
should be a ``deformed Lorentz force'', incorporating a gravitational attraction term.

Again, once the {\em local picture} is taken care of, the {\em global picture}
is the result of applying the homological algebra machinery: 
classical information (external space) as a lattice (chains),
``time'' as cobordisms, quantum information modeled via $SU_2$-coefficients
as internal space, and the emergence of classical properties from quantum ones
via measurement bases, as well as entanglement due to feedback loops,
a consequence off graph (cobordism) extentions and 
internal-external cohomological duality (IE-duality) \cite{I-MPCS}.

The next section is devoted to the fundamental correspondence
{\em Quantum Computing - Special Relativity}, used also as a starting point
in twistor theory.

Section \S 3 investigates Newton's law of the gravitational force as a perturbation of
the electric force.

Since the ``issue'' is not resolved in such a straightforward manner,
alternative ``mathematical guns'' are thrown into the battle.

The concluding section is a preview of the program for developing the new paradigm in physics \cite{I-MPCS}:
QID, and its relations with existing directions of research, notably with String Theory.

The background together with the technical material are confined to the Annex.

\section{Quantum Computing Model of Special Relativity}
There is a correspondence between the Quantum Computing model,
modeling the flow of qubits through the quantum gates of a Quantum Netqork (Q-Net) \cite{QC},
and the Minkowski space $R^{3,1}$ with its Lorentz/Poincare symmetries.

It can be traced to the Klein correspondence with its physical interpretation
in {\em Twistor Theory} \cite{Twistor-Theory}.

It will be denoted by $QC\cong SR$, and hinted to by $2+2\cong 3+1$
(level of dimensions).
It is also referred to as the {\em Hermitian Model} of SR,
since to a 4-vector it associates a hermitian matrix.

\subsection{$SL_2(C)$ and quaternions}
Quaternions, as a division algebra, can be viewed as the result of a double construction,
analogous to the construction of complex numbers:
$$H=C\oplus C^*, J:H\to H, J^2=-1.$$
Here the dual of $C$ has conjugate vector space structure, so $J$ is 
equivalent to a complex conjugate linear map $C\to C$.

It is in fact a {\em generalized complex structure} (GCS) \cite{GCS}.

Conjugation by non-zero quaternions $C:H^{\times}\to Aut(H)$, with kernel $R^{\times}=<1>^{\times}$,
induces the polar decomposition:
\begin{equation}\label{E:SR-CE}
Z(H)=R\to H \to g=(R^3,\times).
\end{equation}
Here $g$ is the {\em angular velocity} Lie algebra with cross-product as a Lie bracket.

The group of unit norm quaternions is $SU_2$.
Under 
Exponentiation of the ad-action:
$$\diagram
g \rto^{Ad} & Aut(g) & x \mapsto ad_x\\
SL_2(C) \rto^{C} & Aut(SU_2) 
\enddiagram$$
yields the action of $SL_2(C)$ on $SU_2$ by conjugation
on itself by rotations $SO_3$  \cite{Kirillov-OM}:
$$ad_\omega (v)=\omega \times v.$$

As a central extension of the Lie algebra $g$,
it is associated to a 2-cocycle related to the Einstein addition of velocities.
It will be called the {\em QC/SR central extension}.

At the level of groups we have the following short exact sequence of 2:1 covering maps:
$$\diagram
SU_2\dto^{2:1} \rto & SL_2(C)\dto^{2:1} \\
SO_3 \rto & SO(3,1)^+.
\enddiagram$$

\subsection{Hermitian model}
It induces a correspondence between hermitan matrices $\xi$ and quadri-vectors
$u=(x,y,z,ct)\in M^{3,1}$ of Minkowski space:
$$
\mathcal{H}\ni \xi=\left(
\begin{array}{cc}
z-ct & x-iy \\
x+iy & z+ct 
\end{array} 
\right).
$$
The conformal structure \cite{Supersymmetry} corresponds to the Minkowski norm:
$$det(\xi)=||u||^2.$$

\subsection{Wick isomorphism}
$sl_2(C)$ has two isomorphic {\em real forms} $sl_2(R)$ 
and $su_2$ (real structure constants):
$$sl_2(R): [K^+,K^-]=S, [K^\pm,S]=\pm S\quad su_2: [J_x,J_y]=J_z\ (cyclic),$$
note also $su_2\cong so_3$ ($\sigma_x$ etc. are Pauli matrices):
$$i \sigma_x \mapsto J_x \ etc.$$
and $su_2\otimes C\cong sl_2(C)$.

A real vector space isomorphism can be defined in terms of canonical bases 
(compare with \cite{Kir-LGLA}, p.47):
$$ WR:sl_2(R)\to su_2, \quad K^\pm \mapsto \sigma_x\pm i\sigma_y, \ S\mapsto \sigma_z,$$
called {\em Wick rotation}.

Then $sl_2(C)$ can be viewed as $g\oplus g^*$,
since $su_2\cong so_3$, with $so_3^*$ the {\em angular momentum}
Lie algebra (dual to $g$).

So $sl_2(C)$ has a canonical generalized complex structure,
determined by the Wick rotation
$$J:sl_2(C)\to sl_2(C).$$ 
The real $Re(\xi)$ of a traceless matrix is the projection on the 
non-compact form $sl_2(R)$, and the imaginary part is the
projection on the compact form $su_2$.

\subsection{The Lorentz 2-cocycle}
The QC/SR central extension determines an infinitesimal deformation
of the Lie algebra $(g, \times)$ with deformation parameter $c$.
Since later we will fully deform $g$,
we view the ``speed of light'' as a {\em formal parameter} $\overline{h}=1/c$
$$u=ct +\vec{v}, \quad u/c=t+\vec{v}\overline{h}+\dots\ .$$
In fact the four-speed of light quadri-vector $\mathbf{c}$ 
consists in the eigenvalues of the Hodge duality isomorphism ($c^2=1/\mu\epsilon$),
and therefore we in fact deform the duality, as we should 
when using the machinery of bialgebra deformation (``quantization'').
The relations between fundamental constants,
interpreted as topological generators (periods):
$$g_m=h/(e/c),\quad g_e=e/c, \quad h=g_m g_e, \quad \alpha=g_m/g_e,$$
together with the duality between the two deformation parameters $\overline{h}=1/c$
and $h$:
$$h\cdot 1/c=\alpha (e/c)^2,$$
will be addressed elsewhere.

Here we only aim to find a possible relation between Newton's gravitational constant
and the above ``fundamental'' ``constants''
(The term ``constants'' refers to the periods,
not to the ``running constants'' which involve the 2-cocycle which makes them variable
with energy, like relativistic mass, electric charge/fine structure constant etc. \cite{Quarks};
the periods are in some sense the ``rest'' values: Lie generators/Hopf primitives).

Returning to the QC/SR central extension of the angular velocity Lie algebra
$$Z(H)\to (H, [,]) \overset{\pi}{\to} (g, \times)$$
where $[q,q']=\pi(q)\times \pi(q')$ since quaternionic product is
$$(t,v) (t',v')=tt'-v \cdot v'+ v \times v'.$$
Note that 3D-vectors are interpreted as velocities 
(elements of the angular velocity Lie algebra $g$), 
rather then position vectors.


Now consider the exponential:
$$exp: {\mathcal{H}} \subset sl_2(C)\to SU_2 \subset SL_2(C).$$
Recall that conjugation by $SL_2(C)$ invaries $SU_2$.

Let $\gamma:g\to H$ be the trivial section $\gamma(v)=v$.

The Einstein velocity addition for parallel velocities \cite{Ungar}
reduces for parallel boosts to an associative and commutative 
group operation:
$$u\oplus v=\frac{u/c+v/c}{1+u/c\cdot v/c}$$
There is an associated 2-cocycle
$$F(u,v)=\gamma(u\oplus v)/(\gamma(u)\gamma(v)).$$
We claim that $\oplus$ is an infinitesimal deformation
of the Lie algebra $g$ addition of velocities
through the logarithm:
$$u\oplus v=u+v+ ...$$
and will be addressed later on.

\subsection{Time and projective space}
``Time'' is a parameter for labeling change.
In mechanics ``change'' occurs in the direction
of ``motion''.
So ``time'' is associated to change in the radial direction
of motion.
A better adapted model of motion is polar decomposition:
$$R^{3x}=S^2\times R_+.$$
With the hidden quantum phase, the role of velocity is that
of de Broglie wave vector $\vec{k}$, i.e. rather an angular velocity
vector (element of $g$).

In QID ``motion'' is a change of angles (direction correlation)
and radial distance (resistence to interaction, or its inverse,
capacity of interaction).

The 2-cocycle of SR corresponds to a central extension
associated to a projective representation.
This in turn comes from a representation at the level of 
projective space as a homogeneous space.
It is the base of the Hopf monopole bundle, the {\em qubit model
of the atom}:
$$S^1\to S^3\to S^2 \quad SU_1\to SU_2\to S^2.$$
In passing we mention that this CS-point of view suggests to
take the Hopf bundles (monopole and instanton) as the ``unifying gauge group''
of all interactions \cite{I-MPCS} ($SU_3$ comes for ``free'').

More precisely $S^2=CP^1$, $MT=Aut(CP^1)=PSL_2(C)$:
$$\xi(z1:z2^*)=(\xi z_1 : \xi^* z_2^*).$$

\section{The Contraction Lorentz to Galilei}
The homogeneous Galilei group is the contraction $c\to \infty$
of the Lorentz group \cite{Kir-LGLA}.

a more general setup is the Anti-de Sitter group
as a 2-parameter deformation of Galilei group \cite{SR21}.

\subsection{Galilei group}
The Galilei group is the Klein group of the 
homogeneous space $R^3$, a semi-direct product
$$ Rot \times Translations$$.
As  a Klein geometry (kinematics) it is better to
view translations as discrete displacements with velocity $v$,
i.e. as an affine space / groupoid:
$${\mathcal{G}}al: \{ A\to B\}.$$
Without spin there is no angular velocity interpretation,
so no (Lie) curvature.

Now abandon the classical and deterministic ``pointwise physics'' 
and adopt a ``bilocal physics'' of interactions (quantum mechanics,
Cramer's tranzactional interpretation,
Feynman-Wheeler theory, Aharonov-Vaidman bistates quantum mechanics etc.).

Translations are banished, every transformation becomes
a two fixed point rotation (Feynman paths: I/O-transformations)
and space-time is ``compactified'' as a category of systems and processes
morphying one system into another (time as cobordisms).

At mathematics level this invites conformal geometry,
and at our local model, Moebus transformations (MT).
Think that such a MT corresponds to two Bloch vectors (four points:
$0, A, \infty, A'$): the ``dumbbell picture'' of QC 
(instanton Hopf bundle).

Wick isomorphism allows to transport the Lie bracket into a cobracket
on $g^*$, yielding a bialgebra structure on $sl_2(C)$.
With such a Poisson-Lie structure, one can apply the quantum double construction.

\subsection{Lorentz force}
It can be viewed as part of Euler-Lagrange equation.

The magnetic term $v\times B$ can be interpreted as a rotation:
$Omega(v)$.

The Euler equation for $G/g$ is \cite{Kir-LGLA}:
$$\dot{\Omega}=ad^*_v \Omega, \quad v\in g.$$

To understand Gravity as a deformation of EM,
consider the whole picture in terms of 
canonical momentum $P=mv+eA$ versus netwon's picture
separate from Lorentz force:
$$d(mv)/dt=q (E + v\times B).$$
Also do not separate particles from fields, as in ``pointwise physics'';
rather consider the categorical interaction picture (``bilocal physics'').
$$d(m_{red} v)=G_N m m'/r^2 + k_e q q'/r^2 + qq' v \times B/q'.$$
here $m_{red}$ is the {\em reduced mass}:
$$2/m_{red} = 1/m + 1/m'.$$
If
$$mc^2=\tau^2, \quad R=log \tau$$
``resistance to change'' adds up in ``parallel''.

This point of view leads to a reinterpretation of Bio-Savart Law as a Gaussian Link;
in this sense, as Descartes was saying, every motion is a rotation ...

So $q,q'$ are indexes (winding numbers/Chern classes of the two Hopf bundles) and 
$m,m'$ are associated to the corresponding momentum maps 
(Hamiltonian circle reduction interpretation
of the qubit as a harmonic oscillator \cite{I-MPCS}).

\section{$SL_2(C)$-Gravity}
We essentially deform $SL_2(C)$-gravity, except we consider Einstein's equations
of General Relativity as expressing a local constraint.
It is the curvature of a reductive Lie group:
$$Ric + Killing/4=0.$$
When ``adding'' homology (Q-net chain / QI cochain cohomology with $SU_2$-coefficients),
one obtains a non-homogeneous version with a topological RHS
which represents matter via 1,2 and 3-periods (fluxons, e/m-charges and spin/action
\cite{Kiehn, Post, EMTC, NLEM}).

So the present goal is to justify Newton's law for Gravity and the associated constant.
The parallel between Newton's constant $G_N$ and Coulomb's constant $k_e$
clearly indicates that this requires unification of internal and external
DOF IE-duality \cite{I-DWT}), which locally is expressed by the 
GCS of $sl_2(C)$ (Wick isomorphism):
$$\frac{\partial L}{\partial \dot{r}}=P=mv+eA \quad \leftrightarrow Wicki:sl_2(R)\to su_2.$$
here $A$ should be the $SU_2$-yang-Mills connection,
as an obvious ``upgrade'' of EM (it should yield ``quark'' DOFs for free;
see also \cite{NLEM}).

\subsection{Relation with Higgs mechanism}
The breaking of symmetry approach to mass \cite{Witten} should be
adapted to the grand unifying Hopf bundle picture
and its associated Gysen sequence.

In view of Stashef's cohomological physics and \cite{I-FL},
it would \underline{not} be a surprise to find that the connecting homomorphism
couples EM and Gravity.
In the homological algebra formulation of QID
interactions should emerge as derived functors of the action
(number of action quanta/entropy and harmonics/fluxons).

\subsection{Quantum gravity}
So we could say that gravity is ``induced'' as a ``quantum perturbation'',
if we compare our approach with the Shacarov's induced gravity.

In fact in QID the Q-net (matter with its classical
properties dependent on a space-time coordinate system) 
emerges dynamically as classical information from
internal QI, under IE-(cohomological) duality.

Gravity does not emerge as in \cite{Verlinde} but rather as a ``non-linear''
(or rather non-commutative) effect; it is the result of deformation via an 
exponentiation process, to be explained next, being in some sense ``quantum''.

\subsection{Mass and electric charge}
We have focused on gravity, but what is electric charge, and why is it different from 
mass as a source of ``forces'' (interactions)?

To address this issue we critically review Newton's law of gravity, with hindsight from
electromagnetism.

First, we should acknowledge that interactions are relational, so if two systems S1 and S2
have parts (subsystems), then if the separation distance is large compared to the internal 
substructure, then then the mutual interaction is well modeled by a bipartite graph:
$$F_{12}=\sum_{ij} F_{ij}\approx N_1 N_2 F,$$
where $F$ is a typical pair of interacting particles.
So in a typical Coulomb law, associated to Newton's law $ma=F$,
LHS will contain a mass coefficient due to system S1,
but in the force term in the RHS will be proportional to the products $M_1M2$ and $q_1q_2$
(we disregard spin at this time):
$$M_1 r''=Coef. f(M1,M2;q1,q2)/r^2.$$
In fact the essence is the potential of an interacting pair of particle:
$$V_{12}=Charge_1 \times Charge_2 \times Conductivity.$$
Here ``space'' conductivity is $1/r$, i.e. distance is ``resistance'' to transfer of momentum-energy
(interactions). The charges are weights corresponding to the types of particles,
via their symmetries (gauge group, Noether theorem).
They are in fact the coupling constants, since the total charges are mathematically speaking
dimensions or indexes (absolute or net charges, like mass and electric charge).

Since we know matter is discrete and mass and charge are additive, we can reduce the global 
low of interaction to the case of a pair of elementary ``particles''.
But there is only one: the qubit / Hopf bundle; the electron corresponds to $S^1$ and proton
corresponds to $S^2$.

Assume that some electrons are ``separated'' from the protons, by having elongated orbitals
shared by the two systems. Let $m^i_\pm$ denote the mass of protons and electrons
in the two systems:
$$m_i=m^i_++m^i_-, \quad q_i=m^i_+-m^i_-.$$
At a fundamental level, if all known forces are different aspects of one interaction,
one cannot distinguish between gravity and electricity, except by an approximation process.
So, in the basic Coulomb law for our two types of interacting particles, 
protons and electrons with masses $m_+$ and $m_-$, let us introduce distinct Coulomb constants
for positive and negative charges:
$$K_+=k+\delta, K_-=-k, \quad m_\pm r''=K_\pm e^2/r^2,$$
where $e$ denotes the MKS unit of electric charge.

Then the force between two neutral H-atoms, i.e. two qubits with preserved $SU_1$-electron symmetry,
is the result of four interaction terms, corresponding to all possible combinations of signs:
$$G=(K_++K_-)^2=\delta^2.$$
We claim that the hypothesis that $G$ is Newton's gravitational constant is worth pursuing,
even in the context of the well known precision experiments which claim that
$e_+=e_-$; experiments are always ``theory dependent'', and sometimes the theory might
suggest the experiment which rules out new effects not derivable from the given theory.

The underlying idea is that ultimately electric charge is a 2-period (integer) \cite{Post,Kiehn,EM},
and the coupling constant is chiral, since the Hopf bundle is.
In other words, gravity is due to parity violation, which in turn, via Dirac's equation
interpreted in terms of QID, is a topological fact about the Hopf bundle.
The ``rest'' is achieved through deformation theory.
At this stage it is not clear if this has to do with the speed of light as a 2nd 
deformation parameter, but it definitely is related to the absence of right handed neutrinos
(see Dirac equation and Weyl spinors).

In conclusion, it is reasonable to investigate if mass is an un-oriented version of charge,
while the electric charge an oriented version, both related to the 2:1 covering 
$SU_2\to SO_3$, derived from the Hopf bundle as a unifying ``gauge group'' and 
concept: the building block of the universe, the hydrogen atom is modeled as the qubit,
the unit of QI, which in turn is the Hopf bundle, with its multiple roles
of harmonic oscillator (Hamiltonian reduction) or local Lorentz symmetries of ``space-time''.
This double function is related to the ``amazing duality'' \cite{Arnold} by conformal inversion
between the harmonic oscillator and Kepler's problem: unification of macro and micro-cosmos
(Einstein equations would appear locally as the fundamental equation of curvature
in a symmetric space $Ricc+Killing/4=0$).

Additional arguments regarding the relation between mass and electric charge will be presented 
later on, in connection with the possible meaning of the vector potential of EM.

\subsection{Chirality and bialgebra deformation}
Now at a technical mathematical level, deformation of EM due to the chirality of the Hopf bundle (Gysen sequence)
is similar to the deformation of universal enveloping algebra of a Lie algebra (quantum groups),
which can be implemented via BCH-formula.
This in turn corresponds to a pull-back of the non-linear group law at the level of the Lie algebra
(generators of symmetry, e.g. Lorentz/Poincare Lie algebra), which corresponds to
the Birkhoff decomposition of the Poisson-Lie group (R-matrix and all that \cite{QG-PC}).
From this angle one can say that the Birkhoff decomposition of $SU_2$-Yang-Mills
yields the gravitational force as a residual Electromagnetic force:
$$F_G=F_{EM}^+-F_{EM}^-.$$
But at the level of the charge multiplicative RHS of Newton's equation, 
the gravitational constant could conceivably be related to the exponential 
of another universal constant, which cannot be other then the fine structure constant:
$$exp(-1/\alpha)\approx 10^{-59}, \quad G\approx 10^{-38}.$$
Looking for additional confirmations, we will briefly speculate on the
meaning of the fundamental constants. We are looking for {\em relations} between
fundamental constants, corresponding to conceptual aspects; we are {\em not} 
looking for {\em numerical coincidences}.

\section{On Fundamental Constants}
By now there is ample evidence that not only action is quantized (Plank-Einstein),
but also magnetic flux and electric charge (fluxons): 
Aharonov-Bohm effect, quantum Hall and Josephson effects,
supraconductivity and Abrikosov lattice (vortex filaments), 
vorticity of superfluid helium.

These quantities were interpreted as periods \cite{Post, Kiehn}:
$$1-periods: \ fluxons \quad h/e,$$
$$2-periods: \ charge \quad e,$$
$$3-periods: \ action \quad h.$$
From the ``mechanical side'' of superfluids \cite{Heliu,London}
another quantum appears: $h/m$.
Since what is really conserved is the canonical momentum $P=mv+qA$,
where $A$ denotes the EM-vector potential, this is no surprise.
In order to understand fundamental constants we need to better understand
the meaning of the vector potential.

\subsection{EM-vector potential as a transfer velocity}
The vector potential (VP) was originally interpreted by Maxwell as momentum per unit of charge
\cite{VP-articles} (see also \cite{OP}).

We go further and interpret it as a transfer velocity, with the following 
justification.

In a BIT the emphasis is on interactions of pair of systems,
{\em without} the usual separation between particles and felds:
$$Pointwise\ physics: \bullet \to, \quad Interaction\ Physics: \bullet\to \bullet.$$
There is an obvious trend towards such a ``bilocal physics'':
Feynman-Wheeler theory, Cramer's transactional interpretation, 
Aharonov-Vaideman two-states quantum mechanics etc.,
not to mention ``categorification of physics'' (Stasheff) and the hidden
compactification in QM via projectivisation.

In order to see the impact on EM, let's rewrite Biot-Savart law in terms of
velocities of the two particles involved ($e_r=r/|r|$):
$$F_B=k q_1q_2 (v_1/r)\times ((v_2/r)\times e_r).$$
the $1/r^2$-coulombian factor should be distributed to both particles,
showing that in some sense the interaction is a {\em linking number}
involving the relative ``angular momentum'' of the two particles:
$$F_B\sim \omega_1\times (\omega_2\times e_r)).$$
This does justice to the Neumann mutual potential as the fundamental 
ingredient of EM as an interaction theory without a background
space-time (and in the spirit of String Theory, if abstracting it from
a particular embedding; so, physics is conformal after all!).

Recalling that the VP is related to current and velocity as in:
$$A=\mu/r \ q v, \quad (continuum: \ J=\rho v),$$
then canonical momentum of a particle $S1$ interacting with a 2nd particle $S2$ becomes:
$$P_12=m_1v_1+\mu/r\ q_1q_2 v_2.$$
Here we have discarded gravity, to focus on the idea of {\em velocity transfer},
as a legitime version of the ``ether dragging'' theory (dark matter etc.).

Recalling Mach's viewpoint of relativity, the VP $A_1\sim v_2$,
which in fact is related to the EM connection 1-form, represents the ``correction''
of the meaning of direction of $v_1$ due to the existence of the 2-nd system;
``space'', as position and direction, is a collective entity due to the presence
of matter, as learned from Einstein's GR.

There are other notable sources supporting this important point: 
1) QFT and collapsing a subgraph: what is then the ``average space''?
2) Cosmology: the Black-hole information paradox.
The solution is the IE-duality \cite{I-DWT}.

So, as a provisional conclusion, interpret $A$ as a transfer velocity
of particle 2 ``correcting'' the ``proper space'' of particle 1; 
of course, modulo the conversion of units to be discussed in the next section.

Notice though, that in view of Descartes point of view
and Maxwell's original point of view \cite{VP-thoughts},
the alternative is to interpret $qA$ (the VP) as a transfered {\em angular momentum}:
$$P_{12}=m_1v_1+\mu q_1q_2 \omega_2, \quad \omega_2=e_r\times (v_2/r),$$
with a typical VP due to the 2nd charge moving with velocity $v_2$ at $r$ given by:
$$A(r)=\mu/r\ v\times e_r, \quad e_r=r/|r|.$$
Regarding IE-duality, cosmic censure-ship etc., there is a need to rewrite
mechanics from its current two tiers: point-mechanics, body-mechanics,
very reasonable hierarchy when accepting the continuum of space and matter,
in to a Mechanics incorporating hierarchy of structure (resolutions), via IE-duality,
which automatically would unite it with thermodynamics, including entropy aspects.

In this vein, a system is modeled as having external parameters mass, charge
and spin (angular momentum), corresponding to internal states ``hidden'' from the
external classical world (C/QI dichotomy), as it was naturally arived at in cosmology
(Kerr black-holes etc.).

Therefore we take as a working hypothesis that mass measures circulation at a 
fundamental level:
$$External\ circulation: \ h/m, \quad Internal\ circulation(fluxon): h/e.$$
This could corresponds to breaking the symmetry of (splitting) the Hopf bundle:
$$S^1\to S^3\to S^2.$$
Think of ``free'' electron and protons as $S^1$ and $S^2$, with their
cohomological 2-period $e$ and 1-period $m$.

This would justify to explore the idea of charge as a 2:1 version of mass ($SU_2\to SO3$),
as we proposed above ($m_\pm$ etc.).

\subsection{The Hodge structure and its deformations}
Bialgebra deformation quantization relies on an ``gluing isomorphism'' (r-matrix),
which is similar to Hodge structure.

To understand better the relation between the continuum approach to physics via
say Lagrangian manifolds and the discrete approach, 
recall the general framework of QID: $SL_2(C)$-cohomology of a dg-coalgebra of 
graph cobordisms (``2-graphs'').
At chain/homology level we have closed and exact cycles (bounded by a ``surface'';
vorticity etc.), and cocycles/cohomology (currents, potentials etc.).

The Hodge structure and decomposition on graphs is the best substitute for
a ``space-time'' decomposition/signature, and at the level of chains is related
cut-cycles decomposition (spanning tree etc.).

At cohomological level it implements the {\em constitutive relations} of say EM:
$$D=\epsilon E, \ H=1/\mu B, \quad \epsilon\mu=1/c^2.$$
It is interpreted by some authors as a ``gauge condition'' \cite{Post, Kiehn},
but it is really an additional structure, reducing the symmetry group. 
It is a compatibility condition similar to the one defining bialgebras
via r-matrices or Frobenius algebras etc.; it is part of the duality.

Let's follow \cite{Post} and consider $\chi:\lambda^2\to \lambda^2$
the middle dimension isomorphism $G=*F$, where $F=dA=(E,B)$ (forms/vectors,
if provided with a metric Hodge duality), and $G=(D,H)$ ($=d^*J$?).
In a diagonal form:
$$\chi=diag[-\epsilon\ I_3, 1/\mu\ I_3].$$
Its invariants are:
$$Hall\ impedence: Z_1=-det(\chi)^{1/3}=\epsilon/\mu,$$
$$Gravity: Z_0=1/3\ Tr(\chi)=1/\mu -\epsilon.$$
The ``units problem'' is ignored for the moment.

Now to keep track of the meaning corresponding to the continuum theories,
recall that in the lagrangian formalism $\partial^2L/\partial p_i\partial p_j$
has the role of metric, or, if looking at the Euler-Lagrange equation, that
of the masses matrix.

In the Hermitian Model (QC$\cong$SR), a qubit/spinor corresponds to a Minkovski 4-vector:
$$det(Q)=||(ct,x,y,z)||.$$
The metric/mass relation as in GR, is related to the constitutive relations (``material''/matter),
say as in:
$$ds^2=\epsilon\ dr^2- 1/\mu\ dt^2, \quad (c^2=1/\epsilon\mu).$$
Of course in GR $c$ is {\em not} and invariant, and in fact in optical physics 
we have 4-light velocities material dependent \cite{Kiehn}.

Hodge structure deformation is reminiscent of GR deformation of metric due to matter.
We will suggest this correspondence via the trademark solution of GR, the
Schwartzschild metric:
$$ds^2=dr^2/(1-r_S/r)-(1-r_S/r)dt^2, \quad r_S=Gm/c^2$$
$$\epsilon=\epsilon_0(1+r_S/r), \mu=\mu_0(1+r_S/r).$$
They are equivalent at first order of approximation.

Now we return to the cohomological relations between periods, which should 
be due to the fundamental unit, the qubit/Hopf bundle:
$$\quad Ext \quad Int$$
$$1-period: h/m \quad h/e$$
$$2-period: m \quad e$$
$$3-period: h.$$
The IE-duality prompts for a ``symmetrization'' of the table, taking
$\overline{h}=1/c$, the 2nd deformation parameter (besides Plank's ``constant'' $h$),
as mass-related/external (deformation of Lorentz/Poincare symmetry group:
Anti-de Sitter group $SO(3,2)$; corresponding in QC with $SU(2,2)$).

The relations between ``fundamental constants'' should reflect the relations
between generators of Chern classes of the cohomology ring of the Hopf bundle
as a coefficient ``group'' for representing the graph cobordism (quiver):
$$h= (h/e) \cdot e, \quad h=h/m \cdot m.$$
One may expect a natural duality between the two deformation parameters:
$$hc=\alpha e^2, \quad \alpha=\frac{h/e}{e/c}=\frac{q_M}{q_E},$$
as for the duality between the quantum group of observables $F(G)$ a la Woronovich
(say $G$ is the Lorentz group) and the quantum group (via bialgebra deformation quantization)
corresponding to the universal enveloping algebra of its Lie algebra:
$$F(G)_c <-> U_h(g).$$
This is conceivable in view of the invariants of the Hodge constitutive map $\chi$.

Note also that quantum Hall effect introduces another ``relation'',
quantization of $h/e^2$:
$$\sigma_{Hall}=\frac{h/e}{e}, \quad \sigma c=1/\alpha.$$
Now recall that on the ``external side'' of mass-space-time we have:
$$Gm_p^2/hc=\beta\approx 10^{-38},$$
and maybe:
$$\ln{G}\sim -1/\alpha,$$
a ``puzzle'' worth staring at for a while.

\subsection{The amazing duality}
The micro and macro-cosmos correspondence via inversion (harmonic oscillator / Kepler problem
\cite{Arnold}) is reminiscent of T-duality from the String Theory \cite{ST-web-introd}.
This is related with the Bohr's planetary model for the H-atom, 
which is essentially the harmonic oscillator.
Now the relation between Bohr's radius (Compton wave length) and 
Schwartzschild radius for H-atom (qubit) is:
$$r_S^p/r_B^p=\frac{Gm_p/c^2}{h/m_pc}=Gm_p^2/hc=\beta.$$ 
Can we relate it with the constitutive relation $\chi$? let's try ...

\subsection{Hodge invariants and gravitational constant}
Recall $P=mv+qA$ or in cgs-units $mv+q/c A$. 
If switching to the bilocal model: $P=m_1v_1+\mu/r q_1q_2 v_2$.
With ``better'' units \cite{Post} and considering unit charges:
$$P_1=m_1v_1+\frac{\mu}{\epsilon} e^2 v_2/r.$$
Now in the bilocal physics angular velocity and angular momentum form
a GCS $g\oplus g^*$, while linear momentum should be translated (compactified
via projective spaces) into angular momentum (``Descartes motto''),
providing additional insight into the distinction between mass and electric charge.

Now assume {\em distance being quantized} as a multiple
of the Bohr radius, with $r/r_B=n$ principal orbital number.
Assuming the non-relativistic Bohr model of H-atom (e.g. \cite{Post-QR}, p.156):
$$E_n=1/2 m_ec^2\ \alpha^2/n^2,$$
the above coupling coefficient $\mu/\epsilon /r$ becomes (with $m=m_e$):
$$Z_{Hall} \frac{e^2}{r_B} \frac{E_n}{\alpha mc^2}
=Z_{Hall} \frac{E_n}{E_\infty},$$
where the rest energy was denoted by $E_\infty$.

From the above relation may speculate that there could be a relation
between $K_\pm$ and $\epsilon$ and $\mu$ of the form
$$K_-=-\sqrt{\epsilon}, \quad K_+=1/\sqrt{\mu}$$
yielding a possible meaning to the other invariant of $\chi$:
$$\sqrt{G}=\delta=K_++K_-=\frac13\ Tr(\chi).$$
The main idea is that, in view of the interpretation of the
EM vector potential, mass and electric charge are directly related,
and therefore so are $k_C$ and $G_N$, Coulomb and Newton's constants.

\subsection{H-atom, protons and electrons}
The relation between EM and Gravity, should be a consequence of the
relation between mass and electric charge.
Their fundamental representatives are the proton and electron, viewed as
``free'' particles, and the neutral H-atom.

Our model is based on the qubit as the fundamental unit 
(QI, spinor, harmonic oscillator, local Minkovski space).
We interpret it as representing the neutral H-atom, with its electron and proton
as the Hopf (monopole) fibration:
$$SU_1\cong S^1\to SU_2\cong S^3\to S^2.$$
The ``free'' instances of the electron and proton are represented by sections,
or equivalently by the direct product $SU_1\oplus SU_2\cong S^1\times S^2$.
Breaking the symmetry could be an alternative mechanism for obtaining 
gravity.

In other words, when viewing an interaction as a link or degree of a bundle 
morphism, as with the ``Neumann'' interpretation of the Biot-Savart law
(see also \cite{NLEM}), then the interaction pairings between 
H-atom $S^3$, proton $S^2$ and electron $S^1$ could yield
 the different coupling coefficients (matrix), which in turn
would reflect into the simplified version $K_+\ne -K_-$.

Note that our interpretation of mass and charge amounts to taking the 
{\em even} and {\em odd} part of a signed scalar quantity (or relative to some other
involution), maybe better with some weights:
$$m=m_+ m_p+m_- m_e, q=m_+ e_p-m_-e_e.$$
The current accepted values include $m_p/m_e\approx 1836$, based on the assumption $e_p=e_e$
within most of the theoretical frameworks (e.g. Bohr's atom or QFT).

Compare with the relativistic mass in terms of rapidity $\theta$:
$$m=m_0 \cosh{\theta}, \quad \tanh{\theta}=\beta=v/c,$$
and ponder whether its counterpart, the electric charge
should vary as the {\em odd part} \cite{Quarks}:
$$q=e \sinh{\theta},$$
towards a generalized rotation of Lorentz/symplectic type.

\section{Velocity as a Gauge Field}
There is enough evidence to interpret
velocity as a gauge field, by exploiting the relation 
with the EM vector potential, especially in the context of a 
background independent theory (BIT).

Recall that the EM-vector potential is thought of as a transfer velocity.
It should be (now) thought of as a transfered {\em relative} velocity,
to gauge away an irrelevant uniform motion (BIT):
$$A\sim v_2-v_1,$$
with $v_i$ values of the velocity gauge field at particle 1 and 2 respectively.

In other words, {\em abandon the coordinate point of view}, as unavoidable in a BIT,
and refer to distances as a ``potential'' via path integration,
possibly multi-valued etc.

The curvature of velocity, or rather momentum and energy, should be interpreted
as acceleration/force related; this at the infinitesimal level.
When exponentiated, the resulting monodromy should be the $SL_2(C)$-matrix 
associated to an exact cycle representing a matter source and acceleration.
It should thought of as the the vorticity of the QI-flow through that cycle:
$$\Omega(cycle)=\int mv+eA <-> sl_2(C), \quad \Psi=U/B(cycle)=exp(i\Omega/\hbar).$$
Here $mv+eA$ should be replaced by the 4-D relativistic version,
corresponding to $sl_2(C)=su_2 \oplus sl_2(R)$ under the Klein correspondence
(QC$\cong$ SR), with the two components: unitary ``space-like'' quantum process
and $SL_2(R)$-Lorentz boost (GR: acceleration=gravity?).

Again we advocate that chirality of Hopf bundle would imply a Birkhoff decomposition 
of the spinor, with the corresponding decomposition of curvature/forces via logarithm,
possible related to gravity (splitting the Hopf fibration):
$$\Psi=\Psi_+ \Psi_-^{-1}, \quad \Omega=\Omega_{EM}-\Omega_G,$$
yielding the ``effective'' coupling constants $K_\pm$ (or the matrix).

\section{Deforming or Quantizing Special Relativity?}
To ``deform'' Lorentz symmetries, we have the following alternative.
The ``winner'' will be the one yielding a gravitational term
when deforming Lorentz force (if any).

\subsection{Deforming quaternions as a Lie algebra: $(1+3)_c$}
Quaternions $\mathbf{Q}$, from the ``1+3'' SR side, 
is a central extension of the angular velocity Lie algebra ($[,]$ the cross-product):
$$R\to {\mathbf{Q}} \to (R^3, \times), \quad =t + (xL_x+yL_y+zL_z)/c,\quad (L_x=i, L_j=j, L_z=k),$$
with the ``speed of light'' $c$ as a central element.

Viewed as a Lie algebra, it can be canonically deformed
via the exponential/logarithm ``non-linear'' transformations,
using BCH-formula. 
Since we think of $1/c$ as a deformation parameter dual to Plank's constant $h$,
we will denote it with $\overline{h}=1/c$ (not to be confused with $h/2\pi$).
In dimensionless formulas $t$ is a number and $|v|/c$ is usually denoted with $\beta$:
$$v\overline{h} = v/c, q=(t,v)=t+v\overline{h}.$$
With these notations BCH-formula is:
$$q\oplus q'=q+q'+1/2[q,q']\overline{h}+1/6([q,[q,q']]-[q',[q',q]])+\dots .$$
The induced deformed velocity addition should be compared with Einstein's 
velocity addition law,which for parallel velocities is:
$$v\oplus_E v'=(v+v')/(1+vv'\overline{h}^2)=(v+v')(1-vv'\overline{h}^2+(vv')^2\overline{h}^4-\dots)$$
$$=v+v'-\{v(vv')+v'(v'v)\}\overline{h}^2+\dots$$
Recall that $a\times (b\times c)=(ac)b-(ab)c$.

This time the deformed addition {\em is} associative for arbitrary velocities, 
unlike Einstein's addition, and yielding a ``Thomas rotation'' time-like term.

\subsection{Quantizing quaternions: $(2+2)_h$}
Since gravity is not velocity dependent, let's try quantizing the quaternion algebra,
viewed from the ``harmonic oscillator'' side:
$${\mathbf{Q}}=(T^*C, \omega).$$
Following the standard procedure for quantizing Heisenberg algebra 
as a central extension of the symplectic space:
$$R\to H\to (T^*,\omega),$$
consider the analog central extension (infinitesimal deformation):
$$R\to {\mathbf{Q}}[h]\to {\mathbf{Q}},\quad
q=z_0|0>+z_1|1>, \quad [[0>,|1>]=ih.$$
together with the corresponding deformation via BCH-formula
(the $UEA(Lie\ algebra)$ side):

On the other hand we have a Moyal product or Weyl quantization on its algebra
of observables (the $F(G)_h$ side).

The resulting Lie algebra/group with one more deformation parameter
should be related with the (anti) de Sitter groups \cite{SR21}.

Now the question is weather the correspondence $QC\cong SR: 2+2=1+3$
is compatible with the duality $F(G)_h^*\cong U(g)_{\overline{h}}$.

\subsection{Deforming the Lorentz force ... again}
In the context of Newton's law,
the Lorentz force $F=q(E+v\times B)$ can be interpreted as coming from a potential,
via the canonical (``electro-mechanic'') momentum:
$$P=mv+qA, V=q(c \phi_E/c-vA), d/dt P=-\nabla V.$$
With electric potential $\Phi_E=1/\epsilon\ q_2/|r|$ in place 
and transfer angular velocity $A=\mu/r\ q_2 v_2\times e_r$ for the EM vector potential,
it becomes:
$$P_{12}=m_1v_1+\mu\ q_1q_2\ \omega_2\times e_r, \quad 
V=q_1 q_2 /|r|\ [\frac1{\epsilon}\ -\mu\ det(r, v_2, v_1)],$$
where the determinant comes from the mixed product $v_1\cdot (r\times v_2)$.

The separation of ``space-time'' components into momentum and 
energy, is performed at the Lie algebra level,
neglecting the non-linearity via exponential.
Will this provide the ``missing'' gravitational term
$m_1m_2 G_N /|r|$?

In fact a relativistic ``potential'' energy formula should involve
a relativistic canonical momentum:
$$P^{rel}=\gamma m_0(c,v)+q(\Phi_E/c, A),$$
$$V^{rel}=P^2/m_0=c^2+q(\Phi_E-vA)+q^2/m_0((\Phi/c)^2-A^2).$$
The 1st term is a constant (discard), the 2nd is the ``old'' potential,
while the 3rd appears as a ``correction'' term.
In fact, in terms of the deformation parameter $\overline{h}=1/c$:
$$(P/\overline{h})^2/m_0=1+(q\overline{h})(\Phi_E-vA)+(q\overline{h})^2 ((\Phi_E/c)^2-A^2)/m_0,$$
is an invitation to a deformation series.

\subsection{... or grading it with external symmetry groups?}
On the other hand a gravitational potential seems to come naturally
together with the different symmetry groups of the various types
of interacting particles'': the ``bound'' H-atom (neutral qubit) 
corresponding to $S^3$, and ``free'' protons and electrons (split qubit) 
with coupling constants corresponding to $S^2$ and $S^1$.
 
The simplified version was discussed in terms of the effective coupling 
constants $K_\pm$ (electric charges slightly not equal, or a parity issue
due to the Hopf bundle).

The electric and magnetic interaction ``force'' would be expressible 
in terms of ``Neumann potentials'' \cite{FromAtoE}
given by linking numbers (see also \cite{Germain, Barrett, EMTC, NLEM}).

The electric force can be represented using a (Gaussian) linking integral
$$\Phi_E=\frac{e_1 e_2}{\epsilon} \int_{S^1}\int_{S^1} \frac{dq\cdot dq_2}{|r|},$$
between two ``time-like'' circle fibers over the space-like charges
(Hopf bundle),
while the ``magnetic force'' (with $\mu$ 
coupling constant in place of $1/\epsilon$),
would result from linking two ``space-time'' circles as ``moving charges''.

As an interesting possibility, 
the higher dimensional topological degrees for linking $S^2$ and $S^3$
could be related with weak and cromodynamics interactions.

Now if the quaternionic product is deformed as above,
there would be corresponding correction terms in the linking integrals.

On the other hand, the author inclines to believe that the gravitational and electric forces
appear to be unrelated due to splitting the $S^3$-Hopf bundle.
At this stage one has to investigate pairs of interacting particles
modeled not just as pairs of qubits, but possibly as the Hopf instanton bundle:
$$\diagram 
S^3\dto \rto^{Q-Gate} & S^3 \dto &  & Internal\ Space:\ S^3\to S^7 \dto \\
S^2 \rto^{Distance} & S^2 & & External\ Space:\ S^4.
\enddiagram$$
The left side corresponds to the discrete picture of QID (quantum networks),
while the right side corresponds to a continuum picture with base a compactified Minkovski
space-time and fiber gauge group $SU_2$.

\section{Conclusions}
Starting from the general framework of QID as a natural framework for 
a Background Space-Time Independent Theory,
the article presents some alternatives to be investigated starting
from the idea that the ``unifying gauge group'' is in fact the Hopf monopole bundle,
underlying quaternion algebra (spinors).
Its roles within the physics interface are:

1) Unit of quantum information (Quantum Computing side: $2+2$), 

2) Harmonic oscillator
(Quantum Mechanics side: $2+2^*$), 

3) Minkovski local space-time ($1+3$).

Lorentz symmetries and Minkowski space seem to describe the infinitesimal/linear picture.
Their deformation into the non-linear/non-commutative realm represents 
an avenue for obtaining gravitational potential as an exponential correction
of the electric force, which together with the magnetic (and possibly weak and strong)
appear as given by linking integrals (topological degrees).

Together with the dynamics of the quantum networks,
as a cohomology theory of dg-coalgebra of graphs, QID seems to provide the 
general mathematical framework and conceptual interface one would expect 
from a ``Theory of Everything''.

\section{Acknowledgments}
The author thanks ISU for the research support,
especially for the sabbatical semester.

The author is also grateful for the excellent research conditions at IHES, where some 
important steps were taken towards the present stage of understanding gravity.
I would also like to thank Maxim Kontsevich and Graeme Segal for stimulating conversations.

\section{Annex: Quantum Gravity - An Open project}
Knowledge accumulates exponentially, and so do physics theories grow in complexity.
This leads to a cumulative development, e.g. the Standard Model as an Expert System 
in particle physics,
similar to the rise of Windows through upgrades and patches.

Some say Physics is in ``trouble'' \cite{Smolin},
and Mathematics has nothing to do with it \cite{Not-even-wrong};
others are less gentle about the lack of a breakthrough in physics as of the
``dark ages of physics'' \cite{CED-mead}.

The author believes that a change in the current R\&D methodology can trigger
a change of paradigm: open math-physics ``programming using a web platform
such as {\em nLab} \cite{JB-nLab}; a new physics model could be designed 
top-down, as an Expert System, taking advantage of the existing math-physics ``toolkits'',
yielding a better and ``cleaner'' alternative to the SM, similar to Linus as 
alternative Operating System alternative to Windows.

in this article the author sketches a more specific plan of attaking the problem
of Quantum Gravity, based on the previous sketches \cite{I-MPCS},
as a follow up of the general principles and incorporated resources from \cite{I-DWT}.

The aim is to formulate a viable Project to be developed according to the above
suggested methodology.

\vspace{.2in}
More specifically, 
a project for implementing Quantum Gravity as a deformation of information dynamics,
as an $SU_2$ ``upgrade'' of Electromagnetism,
to account for the charge-parity violation, is sketched.

It is based on the framework of Quantum Relativity, and the associated 
math-physics tool-box.
The unifying ``gauge group'' being the Hopf bundle, 
underlying the quaternion algebra as a generalized complex structure.
Its quantum group deformation yields Gravity.
The discrete formulation renders the resulting theory already quantum, 
with a particle-wave cohomological duality. 

The new paradigm of Quantum Information Dynamics
is prone for plugging in the main ideas of the Standard Model
regarding the weak and strong interactions.

\subsection{The Main Ideas}       
Gravity is implemented as a deformation of an ``Electromagnetism'',
in the framework of Quantum Information Dynamics \cite{I-MPCS,I-QG}.

The framework is that of a category with duality; for example $SL_2(C)$-graph cobordisms,
representing quantum processes.

The mathematical theory is of the type ``loop groupoid representations'';
it is a Yang-Mills gauge theory bypassing connections by using
Wilson observables directly, corresponding to the connection's monodromy.

In addition to $SL_2(C)$-gravity, we deform the ``gauge group'' $SL_2(C)$ which here is
treated as a bundle enlarging $SU_2$ viewed as the Hopf monopole bundle.
\newcommand{\QQ}{{SL_2(C)_{\overline{h}}}}
The deformation parameter of the resulting quantum group $\QQ$ \cite{HA2QG}
is the reciprocal of the ``speed'' of light $\overline{h}=1/c$ \cite{I-QG}.

We develop the mathematical connection between the Hopf algebra structure of $SL_2(C)$
and Hodge structure on the generalized complex structure \cite{GCS} of the 
deformed quaternions $Q_{\overline{h}}$, containing $\QQ$,
interpreting it from a physical point of view as a deformation of the ``metric''
coefficients representing the ``curvature'' due to matter.
The {\em quantum net} is thought off as a special material with electric and magnetic permittivity
$$\chi=(\epsilon, 1/\mu):\lambda^2\to \Lambda^2 \quad 
\leftrightarrow \quad ds^2=\epsilon\ dr^2-1/\mu\ dt^2,$$
by using an extended version of the Klein correspondence \cite{Twistor-Theory} between Minkowski space:
the Quantum Computing \cite{QC} - Special Relativity correspondence $QC\cong SR: 2+2^*=1+3$.

The {\em constitutive isomorphism} \cite{Post} is an r-matrix entering the deformation of the
Hodge duality structure, and yielding the bialgebra deformation ``quantization'',
or alternatively, the generalized complex structure deformation (Hodge deformation).

Recall that under the $2+2^*=1+3$ above correspondence, the determinant (Kahler form)
is mapped onto the Lorentz metric:
$$
u=(x,y,z,ct)\in (M^{3,1},ds^2) \quad \leftrightarrow \quad
{\mathcal{H}}\ni \xi=\left(
\begin{array}{cc}
a=z-ct & b=x-iy \\
c=x+iy & d=z+ct 
\end{array} 
\right)$$
$$\quad \Rightarrow ||u||=\epsilon\ dr^2-1/\mu\ dt^2=det(\xi)=aa^*-bb^*.$$
Here we take advantage of the grading introduced on the
central extension of the angular velocity Lie algebra ($[,]$ the cross-product):
\begin{equation}\label{E:CE}
R\to {\mathbf{Q}} \to (R^3, \times), \quad =t + (xL_x+yL_y+zL_z)/c,\quad (L_x=i, L_j=j, L_z=k),
\end{equation}
with the ``speed of light'' $c$ as a central element;
the involution is a $Z_2$-graded complex conjugation: 
$$*(z+ct)=z-ct, \quad *(x+iy)=x-iy.$$

\subsection{Charge-conjugation violation}
The classical two charges $\pm$ are the eigenvalues of the involution $*$.
The quantum deformation yields a {\em twist} $*^2=\delta$ \cite{QG-PC},
responsible for breaking the symmetry.
In this way we implement the idea that Gravity is due to a charge-conjugation violation,
which produces an additional term in Newton's law with Coulomb's potential for electric force
(static case).

in a Lagrangian formulation, with the corresponding Euler-Lagrange equation, 
the deformed symmetries yield via Noether Theorem a quantum charge which 
does not obey the standard Fermi statistics, due to a deformed braiding 
$$\sigma=R\circ Twist, \quad R=Rota-Baxter\ operator;$$
for details on the relation between r-matrix, bialgebra deformation 
(Poisson-Lie groups, left/right dressing action etc.), Yang-baxter equation,
braiding etc. see \cite{QG-PC}.

Deforming $SL_2(C)$, or rather Clifford algebra of quaternions, yields 
the quantum determinant which is interpreted in \cite{I-QG} as the Newton-Coulomb 
coupling constant:
\begin{equation}\label{E:NC}
Qdet
\left(
\begin{array}{cc}
q & im'\\
-im& q' 
\end{array} 
\right)=qq'-e^{-1/\alpha}\ m m'.
\end{equation} 

\subsection{The Category with Duality and CPT}
The category with duality corresponding to the quantum group implements
mathematically the Charge-Parity-Time Theorem, as mentioned in \cite{I-arrow}.

The twist / braiding which breaks the charge-conjugation symmetry
also corresponds to the chirality of the ``gauge group'', i.e.
the chirality of the Hopf monopole bundle.
In a precise sense QID posseses not only electric and information charges (qubits),
but also {\em magnetic charges} as degrees of the hopf bundles ``sitting'' over the
nodes of the Q-net.

The interaction picture model consists of a pair of quabits mapped to the 
Hopf instanton bundle $S^3\to S^7\to S^4$, again a ``topological
fragment'' of the algebraic double $SU(2,2)$.

\subsection{Dirac's equation}
In the categorical framework, the splitting of the Dirac spinor into the chiral
Weyl spinors obeing Dirac's equation \cite{Dirac-Weyl} is essentially
a Birkhoff decomposition of the representation:
$$\Psi=\Phi^+/\Phi^-, \quad D=d+d^*$$
for details see \cite{Spitzer} and related articles.

So the deformation of $SL_2(C)$ induces a parity violation also 
(see Hermitian categories \cite{Bakalov}),
which allows to account for neutrinos as twist, and the lack of left-handed ones.

The quarks and quark charges appear from the interplay between $SU_2$ and 
$SO(3,C)$, with the corresponding Hopf degrees (monopole and instanton bundles:
see the link number interaction model).

\subsection{The Fundamental Constants}
The two approximate eigenvalues $e^\pm\approx \pm$ of $*$, 
i.e. the two charges will obey a constraint coming from the central extension
of the angular velocity Lie algebra:
$$hc = e^+e^-.$$
here $h$ is Plank's constant, the trademark of the angular velocity Lie algebra
$g=(R^3,\times)$, $c=1/\overline{h}$ is the ``speed'' of light 
related to the deformation parameter (central element) $\overline{h}$.

The constraint satisfied by the deformed involution/antipode of the quantum group 
is:
$$c^2=\frac1{\epsilon\mu}, \quad \overline{h}^2=\epsilon / (1/\mu).$$
The other ``quantizations'' of magnetic flux, hall conductivity
are related to the invariants of the deformed Hodge structure $\chi$:
$$Quantum\ flux: g_M=hc/e, \quad Quantum\ Charges: e^+=e-\delta, e^-=-e-\delta,$$
$$Hall\ conductivity=\frac{\epsilon}{\mu}=det(\chi)^{1/3}.$$
Recall that $\delta$ yields the Newton's gravity term as a correction to 
Coulomb's law (\ref{E:NC}).

The fine structure constant appears as 
an asymptotic ratio, with the classical approximation for charges
$g_E=e$ being used:
$$\alpha=e^2/hc=g_E/g_M;$$
for details regarding the interplay of $\epsilon, \mu, \chi, c$ see \cite{Post-constitutive}.

Returning to Dirac equation, as Dirac stated \cite{Jehle}, p.3,
electric charge should be derived from $h$ as a square root; that is what we
claim above.
Splitting the extension yields a double cover and $e=\pm \sqrt{hc}$;
in the deformed case, the chirality should be included.
The corresponding covering map should be responsible for the ``average''
yielding the fine structure constant, essentially a zeta value \cite{I-arrow}
combined with a Casimir element/ central charge
\begin{equation}\label{FSC} 
\alpha=e^2/hc \quad \leftrightarrow \quad ``137''\ h= Det(Q)\ \overline{h}, \quad Det(Q)=e^+e^-.
\end{equation}

Now on the ``mass'' side of the canonic momentum $P=mv+qA$, which is implemented
in the context of the generalized complex structure $TC\oplus T^*C$ on $\QQ$,
the deformation of charges yields a gravitational constant
$$G_N=e^{-\beta}, \beta \leftrightarrow 1/\alpha.$$
The relation should bypass $\alpha$, being expressed in terms of $h,c,e^\pm$.

It seems that the charge defect $\delta$ \cite{I-QG}
is related to the other invariant of the deformation $*$:
$$G_N=\delta^2 \leftrightarrow Tr(\chi)=\epsilon - 1/\mu, \quad G_N/k_C=e^{-\beta}.$$
The numeric factors were discarded for emphasis of the conceptual
meaning, in the sense of dimensional analysis.

\subsection{Harmonic analysis on groupoids}
So a quantum process is modeled as a representation of a quantum network,
with coefficients in a Hopf algebra (quantum group $\QQ$).

Now in analogy with Fourier analysis,
the spectra of graph cobordisms \cite{Alg-GT, I-FL} should yield the
``vibration modes'' of the Q-net.
From the Riemann surfaces side of the picture via ribbon graphs
and Turaev-Feynman calculus \cite{Turaev},
the modes should correspond to integral periods of the Hodge structure.

From the physical point of view, the ``frequencies'' are due to 
a resonance condition:
$$\omega^2=1/LC \leftrightarrow c^2=1/\epsilon \mu.$$
in a direct analogy with electric circuits as $SU_1$-voltage graphs
(circle bundles over chain complexes, Pontryagin duality etc.),
inductance corresponds to mass, and ``elasticity'' of the
harmonic oscillator corresponds to capacitance (parameter of the
qubit or quantum register viewed as a harmonic oscillator \cite{I-QG}).

\subsection{Relation to the Standard Model}
The relation between the ``gauge group of the SM $SU_1\times SU_2\times SU_3$
and the Hopf bundles, monopole and instanton,
underlying the main object, the symplectic qubit $X=T^*H$ (reality is One dimensional internally,
and potentially infinite dimensional externally), and its processes $G=Aut(T^*H)$,
makes the transfer of ``technology'' to the new paradigm of QID/QG possible.

As a previous attempt to implement quarks and mesons using links,
we mention \cite{Jehle-particles}.
This attempt was done in the spirit of String Theory, 
embedding circles in a configuration manifold,
without being relativistic.
These two important ``lessons'' can be used to implement
the quark model and strong interaction in QID, using {\em linking numbers} \cite{I-QG}
(homological algebra invariants and derived functors; dg-coalgebras of graph cobordisms
and its equivariant cohomology with coefficients in $\QQ$ \cite{I-CFG}).

\subsection{Further Developments}
The Quantum Computing model of Special Relativity, once deformed,
seems to be the ideal ground for developing Quantum Gravity as a
Grand Unified Theory (Theory of Everything).

The Hodge periods/Quantum Groups approach is prone for an
abstract formulation in terms of hermitian categories;
a Tanaka-Krein duality should be related with the Connes-Kreimer 
Hopf algebra approach to renormalization \cite{CK}, where Rota-Baxter Algebra
with RB-operator $R$ is a different face of an r-matrix 
and $dd^*$-structure leading to deformation \cite{I-HREN, I-LTDT}.

\vspace{.2in}
A web-based platform such as John Baez's nLab \cite{JB-nLab}
seems to be ideal for the development of this project.


\end{document}